\documentclass[prl,aps,epsfig,twocolumn,nofootinbib]{revtex4}

\usepackage{amsmath}
\usepackage{amsfonts}
\usepackage{amssymb}
\usepackage{epsfig}
\usepackage{color}
\usepackage{ulem}
\usepackage{float}
\usepackage{mathrsfs}
\usepackage{indentfirst}
\usepackage{ifpdf}
\ifpdf   
  \RequirePackage[pdftex]{hyperref}
\else    
    \RequirePackage[dvips]{hyperref}
  \fi
\hypersetup{
       bookmarksnumbered,%
              colorlinks,%
               linkcolor=blue,%
               citecolor=blue,%
              plainpages=false,%
            pdfstartview=FitH}

\definecolor{dyellow}{rgb}{1.,0.8,.0}
\definecolor{myblue}{rgb}{.1,.1,.7}
\definecolor{dcyan}{rgb}{.0,.6,.6}
\definecolor{dmagenta}{rgb}{0.6,0.0,0.6}
\definecolor{brown}{rgb}{0.6,0.2,0.}
\definecolor{darkblue}{rgb}{.0,.0,0.5}
\definecolor{darkred}{rgb}{0.75,0.0,0.0}
\definecolor{orange}{rgb}{1.,.6,.0}
\definecolor{dorange}{rgb}{0.8,.4,.0}
\definecolor{darkgreen}{rgb}{0.0,0.6,0.0}
\definecolor{purple}{rgb}{.4,.0,.4}
\definecolor{lightgrey}{rgb}{0.7,0.7,0.7}

\def\od{{\rm d}}
\def\bc{\begin{center}}
\def\ec{\end{center}}
\def\be{\begin{eqnarray}}
\def\ee{\end{eqnarray}}

\newcommand{\omits}[1]{}

\newlength\CJKtwospaces
\omits{}

\begin{document}
\title{Propagation effect of gravitational wave on detector response}

\author{Zhe Chang, Chao-Guang Huang, and Zhi-Chao Zhao}

\affiliation{Institute of High Energy Physics and Theoretical Physics Center for
Science Facilities, \\ Chinese Academy of Sciences, Beijing 100049, China}

\begin{abstract}
The response of a detector to gravitational wave is a function of frequency.
When the time a photon moving around in the Fabry-Perot cavities is the same
order of the period of a gravitational wave, the phase-difference due to the
gravitational wave  should be an integral along the path.  We present a
formula description for detector response to gravitational wave with varied
frequencies. The LIGO data for GW150914 and GW 151226 are reexamined in this
framework.  For GW150924, the traveling time of a photon in the LIGO detector
is just a bit larger than a half period of the highest frequency of
gravitational wave and the similar result is obtained with LIGO and Virgo
collaborations. However, we are not always so luck. In the case of
GW151226, the time of a photon traveling in the detector is larger than
the period of the highest frequency of gravitational wave and
the announced signal cannot match well the template with the initial
black hole masses 14.2M$_\odot$ and 7.5M$_\odot$.
\end{abstract}

\pacs{04.30.Nk 
04.80.Nn 
}

\maketitle

The detection of gravitational waves (GW) in the event GW150914 by the two advanced
detectors of the Laser Interferometer Gravitational-wave Observatory
(LIGO) \cite{LVC00} opens a new era for the direct
detection of GW \cite{LVC01}, searching black hole coalescence \cite{LVC02}
and `heavy' black holes with more than 25 solar mass \cite{LVC03},
test of general relativity \cite{LVC04}, understanding the astrophysical
environment of black hole formation \cite{LVC06}, etc.  In one words,
the era of multi-messenger astronomy has begun \cite{{LVC00},{LVC01},{LVC02},{LVC03},{LVC04},{LVC06},{Blair},{Lee},{Fan}}.

The analysis of GW150914 shows that the initial black hole masses are 36M$_\odot$ and 29M$_\odot$, which is heavier than the previous known
stellar-mass black holes \cite{StellarBHmass}.  In the newly announced black
hole merge event, GW151226 \cite{LVC10}, the initial black hole masses are about 14M$_\odot$ and 8M$_\odot$, which fall into the known mass range of
stellar black holes in the previous observations.  The signals
from noise for GW150914 and GW151226 are extracted by the same methods
(for example, see \cite{{Black},{MatchFilter}}).  It seems to make the observational results more reliable.

In this Letter, we try to provide a different method from Ref. \cite{Black} to determine
the detector response to GW.  Our
basic observation is that a photon moves along null geodesics in Fabry-Perot
cavities (except the reflections by the mirrors) no matter whether GW exists
or not.  We shall present a frequency-dependent modulation for detector
response to GW, and reexamine the GW150914 and GW151226 data
carefully.

For simplicity, we consider a GW with frequency $f$ and polarization
$h_{+}$ incident normally to a detector, and assume the polarization
along the detector's $x-y$ arms.  The lengths of the two cavities are
supposed to be the same.

\begin{figure}[b]
\begin{center}
\includegraphics[width=7.6cm]{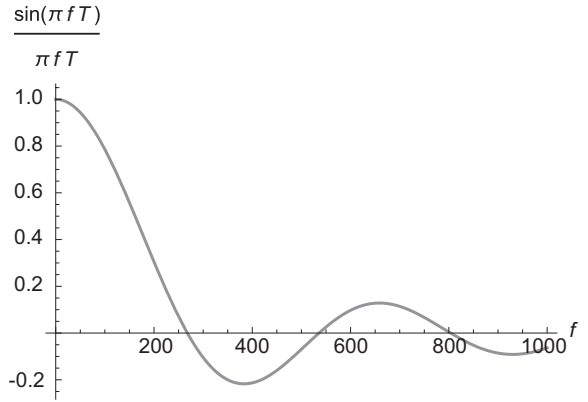}
\caption{Modulation function due to the propagation effect of GW.}
\end{center}
\end{figure}
The metric of the linearized plane GW has the well-known form
\begin{equation}
\od s^2 =- \od t^2+(1+h_{11})\od x^2 + (1-h_{11})\od y^2+\od z^2 \, .
\end{equation}
The time-difference that a photon travels for a round trip
along null geodesics in two Fabry-Perot cavities in a detector is
\begin{eqnarray}
\Delta t_1 &:=&\frac{2}{c}\left (\oint
(1+\frac{1}{2}h_{11})\od x -\oint
(1-\frac{1}{2}h_{11})\od y\right ) \notag \\
&=&\frac{1}{c}\left(\int_0^{L}h_{11}\od x-\int_{L}^0h_{11}\od x\right .\notag \\
&& \qquad \left .+\int_0^{L}h_{11}\od y -\int^0_{L}h_{11}\od y \right) \notag \\
&=& \frac{4L}{c}\frac{\sin(2\pi f L/c)}{2\pi f L/c} h_{11}(t+L/c)\mid_{z=0}, \label{eq:1rt}
\end{eqnarray}
where $L$ is the length of a Fabry-Perot cavity, $t$ is the initial time of the
round trip, $z=0$ denotes the plane the detector lies on.  It is the double of the time-difference in a same-size Michelson
interferometer \cite{Black}.  Since the frequency of GW observed is low enough
($f<700$Hz), the inequality $2f L/c \ll 1$ is always valid, and
Eq.(\ref{eq:1rt}) reduces to
\begin{equation}
  \Delta t_1(t)=\frac{4L}{c}h_{11}(t+L/c)\mid_{z=0}.
\end{equation}
\begin{figure}[H]
\begin{center}
\includegraphics[width=7.6cm]{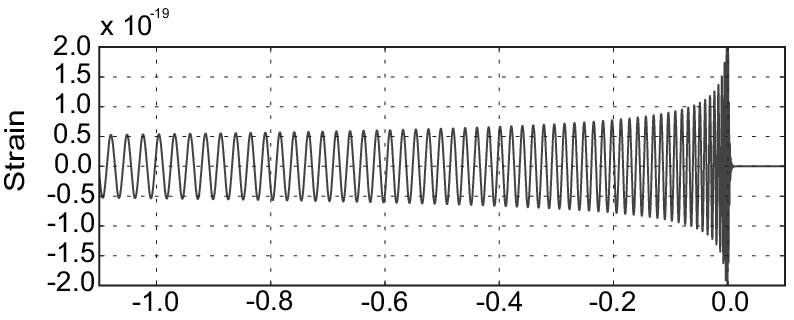}
\includegraphics[width=7.6cm]{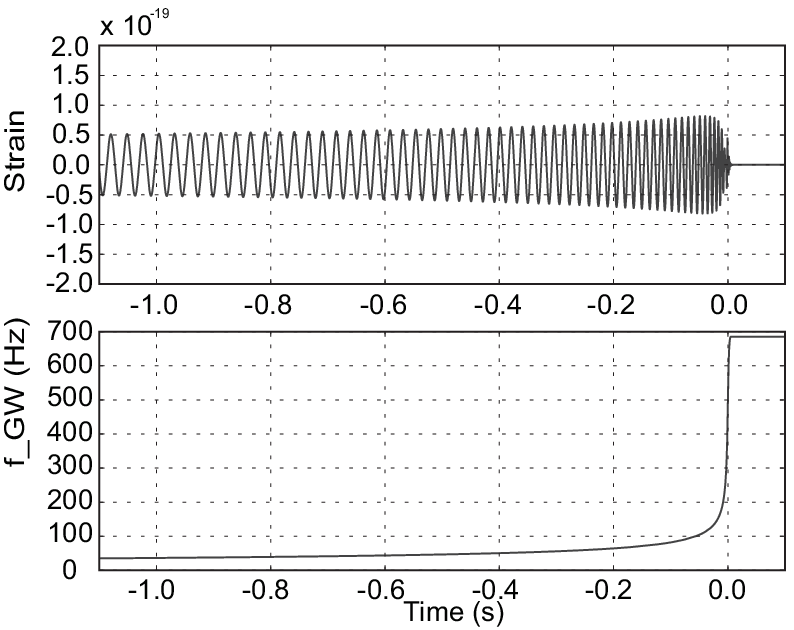}
\caption{The first panel: Strain of time of the template for plus polarization
for GW151226 provided on LIGO website.  The second panel: Strain of time of
the same template with the propagation effect of GW being taken into account. The third
panel: The varying frequency of GW with time. }
\end{center}
\begin{center}
\includegraphics[width=7.6cm]{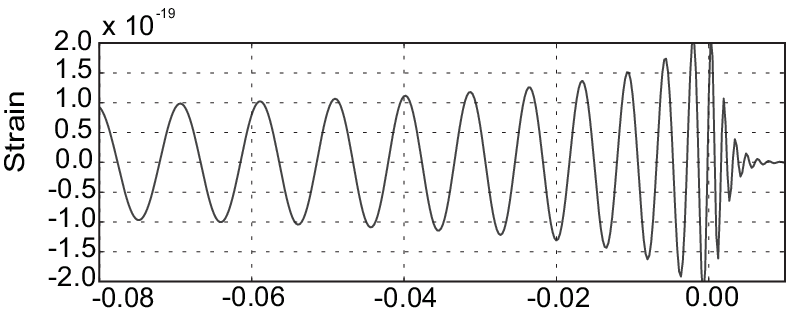}
\includegraphics[width=7.6cm]{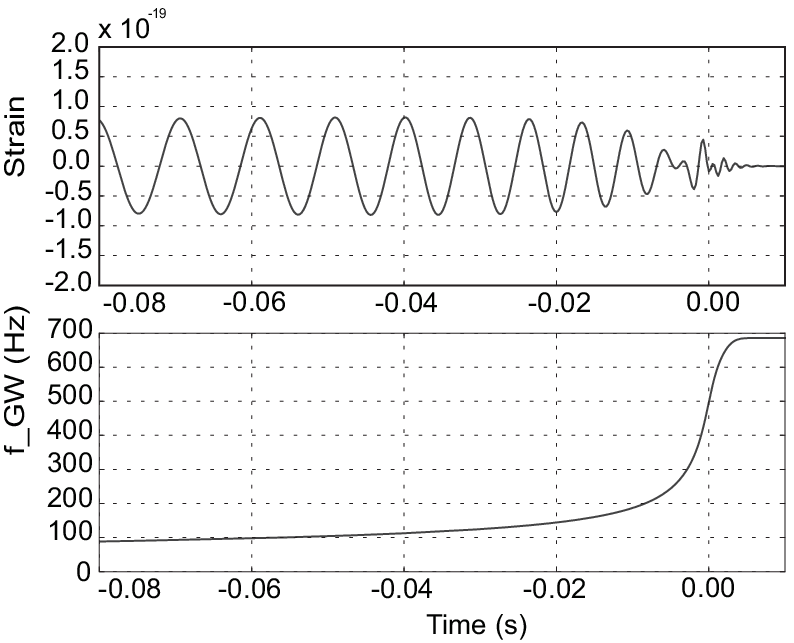}
\caption{The high-frequency part of FIG. 2.}
\end{center}
\end{figure}
\noindent
In average, each photon will
travel $\langle N\rangle \approx {\cal F}/\pi \approx 140$ round trips in the cavity,
where ${\cal F}$ is the finesse
of the cavities.  Due to the propagation of GW,
the gravitational field in which the photon travels in cavities are different time by time.
Then, the time-difference that a photon travels for  $\langle N\rangle$-round trip should be
\begin{figure}[t]
\begin{center}
\includegraphics[width=7.6cm]{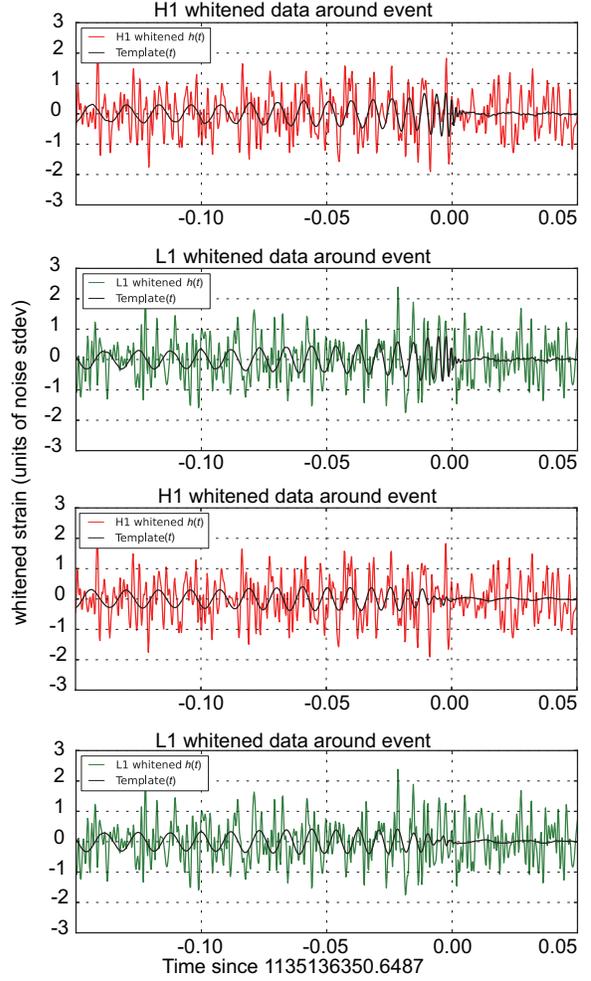}
\caption{The first and second panels show the matching of the
GW signal GW151226 observed by LIGO and the template
without taking into account the propagation effect of GW.
The third and fourth panels show matching of the GW signal
GW151226 observed by LIGO and template by taking into
account the propagation effect of GW. The red and green
curves denote GW strain projected onto each detector
of the LIGO, and the black curve is the template.  }
\end{center}
\end{figure}
\begin{eqnarray}
\Delta t_{\langle N\rangle}&=&\frac{4L}{c}\sum_{n=1}^{\langle N\rangle}h_{11}\mid_{z=0} \notag\\
&=&\frac{4L}{c}\frac {\cal F}{\pi}\frac{\sin(\pi f T)}{\pi f T}h_{11}(t+T/2)\mid_{z=0},
\end{eqnarray}
where $T= 2L\langle N\rangle/c$. The signal recorded by a detector is proportional to the
time-difference.  In the case of $2 L f \langle N\rangle/c \ll 1$, the wave length of GW is much
longer than the $2\langle N\rangle$ times of cavity length,
\begin{equation}
\Delta t_{\langle N\rangle}\approx \frac{4L}{c}\frac {\cal F}{\pi} h_{11}(t+T/2)\mid_{z=0}~.
\end{equation}
\noindent
The signal will be enhanced by $2{\cal F}/{\pi}$ times compared with a same-size
Michelson interferometer.  However, in a generic case,
the measured signal should have a suppression factor $\sin(\pi f
T)/(\pi f T)$, which is a function of the frequency of GW.
Figure 1 shows the propagation effect of GW on the signals measured
by a detector.  The effect makes the signal vanish when the photon's traveling time equals
the inverse of GW frequency.  From FIG. 1, one can find that
even when the frequency of GW is about 150 Hz (at which the
gravitational radiation is strongest in the event GW150914), the suppression factor should be about 0.558.

In the following, we reexamine the GW150914 and GW151226 by taking
into account the propagation effect of GW on the signals detected by the
LIGO detectors.  The program, the data, and the best fit template used here are
all provided by LIGO\footnote{https:$\backslash\backslash$losc.ligo.org$\backslash$tutorials$\backslash$\label{ft:website}}.
It should be noticed that the results obtained here
will not match precisely with what released by the LIGO and Virgo
collaborations, due to various subtleties in the analysis that are
not addressed on LIGO website$^{1}$ and thus are ignored  here.
But these various subtleties are smaller than the
propagation effect of GW considered here.

Figures 2 and 3 show the comparison between the template for GW151226 provided
by LIGO$^1$  and the one in which the propagation effect of GW has been taken
into account. Figure 3 is the high-frequency part of FIG.2.
It is easy to notice that for the low-frequency of GW the propagation effect is small.

Figure 4 is the matching of GW signal and the templates.
When the propagation effect of GW is ignored, the result is similar to LIGO
and Virgo collaborations.  It can be read out from first two panels of
FIG. 4. The third and fourth panel show that, after the propagation
effect of GW is taken into account, it is difficult to ensure the matching
of the signal and template.

We also reexamined the data for GW150914. The results are presented in
Figures 5, 6, and 7.
Due to the frequency  of the gravitational radiation in the event GW150914
is much lower than GW151226, the
propagation effect for GW150914 is smaller than that for GW151226.
Even so, there is still an observable propagation effect of GW in high-frequency part.
\begin{figure}[H]
\begin{center}
\includegraphics[width=7.6cm]{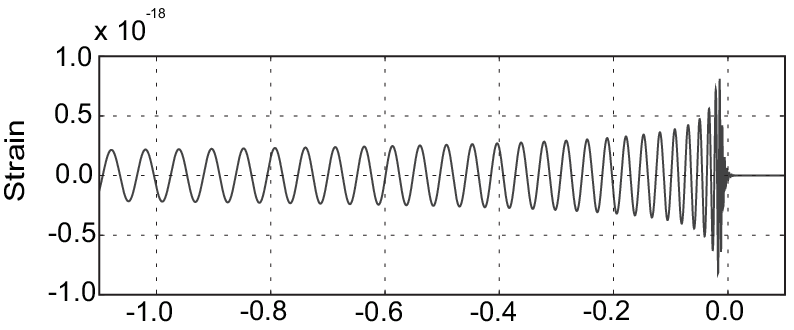}
\includegraphics[width=7.6cm]{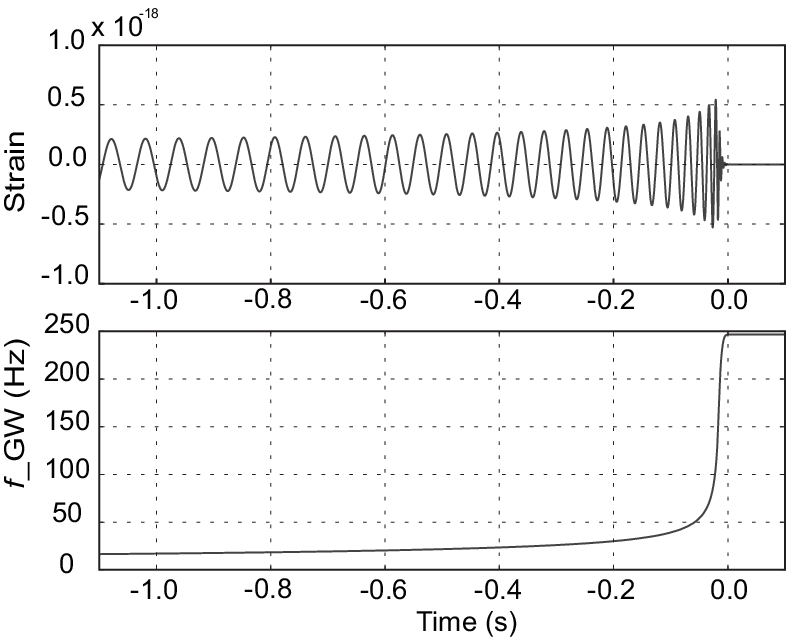}
\caption{The first panel: Strain of time of the template for plus polarization
for GW150914 provided on LIGO website$^1$.  The second panel: Strain of time of
the same template with the propagation effect of GW being taken into account. The third
panel: The varying frequency of GW with time.}\bigskip
\end{center}
\begin{center}
\includegraphics[width=7.6cm]{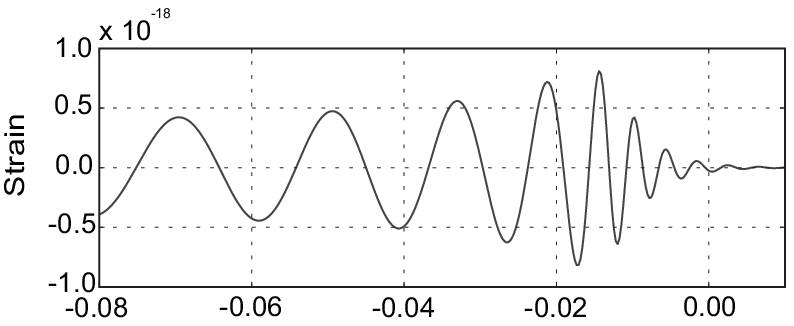}
\includegraphics[width=7.6cm]{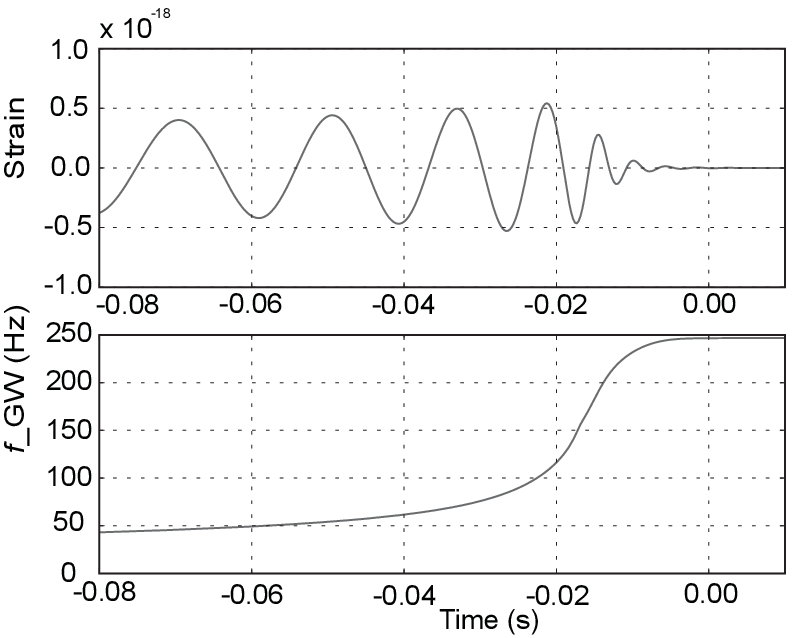}
\caption{The high-frequency part of FIG.5.}
\end{center}
\end{figure}
As a conclusion, the propagation effect for high-

\begin{figure}[t]
\begin{center}
\includegraphics[width=7.6cm]{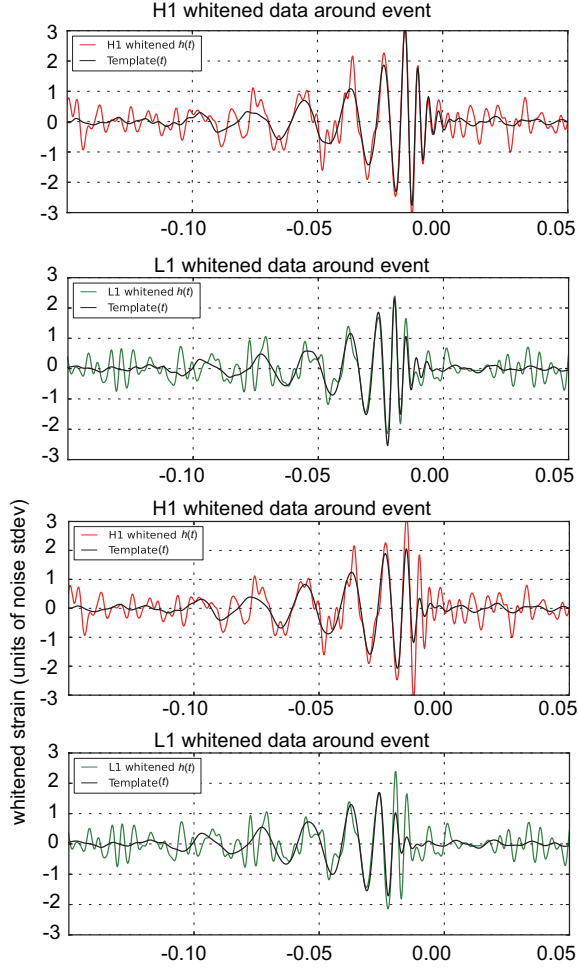}
\caption{The first and second panels show the matching of the
GW signal GW150914 observed by LIGO and the template
without taking into account the propagation effect of GW.
The third and fourth panels show matching of the GW signal
GW150914 observed by LIGO and template by taken into
account of the propagation effect of GW. The red and green
curves denote GW strain projected onto each detector
of the LIGO, and the black curve is the template.}
\end{center}
\end{figure}

\noindent
frequency GW is important in
matching the signals observed with templates.  Although our analysis disfavors GW151226,
it is still possible to find more reliable GW signals from observation data according
to the new templates after the propagation effect is taken into account.  Finally,
it should be remarked that
the modulation function for detector response, $\frac{\sin(\pi f T)}{\pi f T}$,
is calculated from GW with a fixed frequency and amplitude.  The frequency and amplitude
of the actual GW signals vary time by time.  So, the quantitative analysis should be an
integral of the time-difference of a photon along the paths traveling in the two cavities.

\bigskip

\section*{Acknowledgment}

We would like to thank Profs. Yong-Gui Li, Zhou-Jian Cao, and Xi-Long Fan for helpful discussion. This work is supported by National Natural Science Foundation of China under the grants
11275207 and 11375203.

\end{document}